\documentclass{aa}  
\usepackage{graphicx}
\usepackage{txfonts}
\begin{document}
   \title{ High and low states of the system AM Herculis }

   \subtitle{ }

   \author{Kinwah Wu
          \inst{1,2}
          \and
          L\'aszl\'o L. Kiss \inst{2}
          }

   \offprints{K. Wu}

   \institute{Mullard Space Science Laboratory, 
                      University College London, 
                      Holmbury St Mary, Surrey RH5 6NT, United Kingdom \\
              \email{kw@mssl.ucl.ac.uk}
         \and
          School of Physics A28, University of Sydney, NSW 2006, Australia \\
             \email{l.kiss@physics.usyd.edu.au}
             }

   \date{Received ...; accepted ...}

 
  \abstract
   { We investigate the distribution of optically high and low states  
           of the system AM Herculis (AM Her).  
          }
   {  We determine the state duty cycles,   
            and their relationships with the mass transfer process and binary orbital evolution 
            of the system. 
          }
   {  We make use of the photographic plate archive of the Harvard College Observatory 
          between 1890 and 1953 
          and visual observations collected by the American Association of Variable Star Observers 
          between 1978 and 2005.  
      We determine the statistical probability of the two states, their distribution and recurrence behaviors. 
         }
   {  We find that the fractional high state duty cycle of  the system AM Her is {63\%}. 
      The data show no preference of  timescales  on which high or low states occur.   
       However, there appears to be a pattern of long and short duty cycle alternation, 
           suggesting that the state transitions retain memories.  
       We assess models for the high/low states for polars (AM Her type systems).  
       We propose that the white-dwarf magnetic field plays a key role 
           in regulating the mass transfer rate and hence the high/low brightness states, 
           due to variations in the magnetic-field configuration in the system.         }
   {}


   \maketitle
%

\section{Introduction} 

Magnetic cataclysmic variables (mCVs) are close interacting binaries  
   containing a Roche-lobe filling low-mass companion star, 
   usually an M dwarf, transferring material  to a magnetic white dwarf.  
Most mCVs have orbital periods around $1.3 - 5$ hours, 
   and the magnetic fields of the white dwarfs in mCVs 
   are sufficiently strong ($B \sim 10^6 - 10^8$~G) 
   that the accretion flow is field-channelled near the white dwarf. 
mCVs are strong X-ray emitters, 
   and many mCVs were first identified in X-ray surveys.   

The two main subclasses of mCVs are the polars and the intermediate polars.   
Polars are also known as AM Herculis type systems. 
Unlike other cataclysmic variables  and binary X-ray sources,    
   polars do not have an accretion disk.  
All components in the system are locked by a strong white-dwarf magnetic field  
   into synchronized  (or almost synchronized) rotation.  
Intermediate polars contains white dwarfs with weaker magnetic fields.   
An intermediate polar has an accretion disk, 
    but the inner disk is truncated by the white-dwarf magnetic field.  
White dwarfs in intermediate polars are often fast-spinning, 
   with a spin rate roughly 10 times higher than the rate of the orbital rotation.   
For a review of cataclysmic variables and polars, 
   see Cropper (1990); Warner (1995); Wu (2000); Beuermann (2002); Wu et al. (2003).  

Polars are found to show alternating bright and faint optical/X-ray states,    
   commonly referred to as high and low states, respectively.   
In the absence of an accretion disk,  
   the high and low states of polars 
   are likely to be consequences of changes in the mass-transfer rate in the system.  
It has been argued 
  that the high/low states of polars are related to  
   the atmospheric magnetic activity of the mass-donor star. 
In the star-spot model (Livio \& Pringle 1994),   
   a low state is caused by a temporary cessation of mass transfer, 
   due to a magnetic  spot on surface of the Roche-lobe filling mass-donor star 
   traversing  the inner Lagrange  (L1) point.  
While we may have some understanding 
   of the long-term variations in the mass transfer of  cataclysmic variables,  
   we are still uncertain what causes the high and low states and the state transitions of polars.     
 
The bright polar AM Herculis (AM Her =  4U~1813+50, 
   with a visual magnitude $V \sim 13 - 15.5$  and an orbital period of 185.6~min)   
   has been monitored in the optical bands for more than 100 years.   
The optical brightness of the source has shown large variations.     
 An early study (Feigelson et al.\ 1978)  
   claimed that  there was no strong evidence of a well defined low state.     
A later study (G\"otz 1993), however, showed the presence of high and low states 
   with good statistical significance.  
On the other hand, 
   analysis of the fraction of polars in high and low states 
   during the {\it XMM-Newton} and {\it ROSAT} X-ray survey observations (Ramsay et al.\ 2004)     
   suggested that polars spend roughly half of their time in a high state  and half in a low state.   
While surveys of ensembles of systems might establish the effective 
   duration of the high/low state duty cycles, 
   to determine the characteristics of the state transitions and the driving mechanisms   
   we need to analyze the long-term behavior of individual systems.  
  
In this paper we present a quantitative analysis of  high/low states and their transition 
   for the polar AM Her.  
We use the photometric data of the system taken in the last 115 years.  
We assess the models for the state transitions and propose a possible scenario for the process.  
We organize the paper as follows. 
In section 2 we present the data and analysis; in section 3 we comment on the statistics 
    and in section we discuss the magnetic spot model and an alternative. 
A brief summary is given in section 5.


\section{Data sources and analysis}


\begin{figure*}
\begin{center}  
\leavevmode
\includegraphics[width=\textwidth]{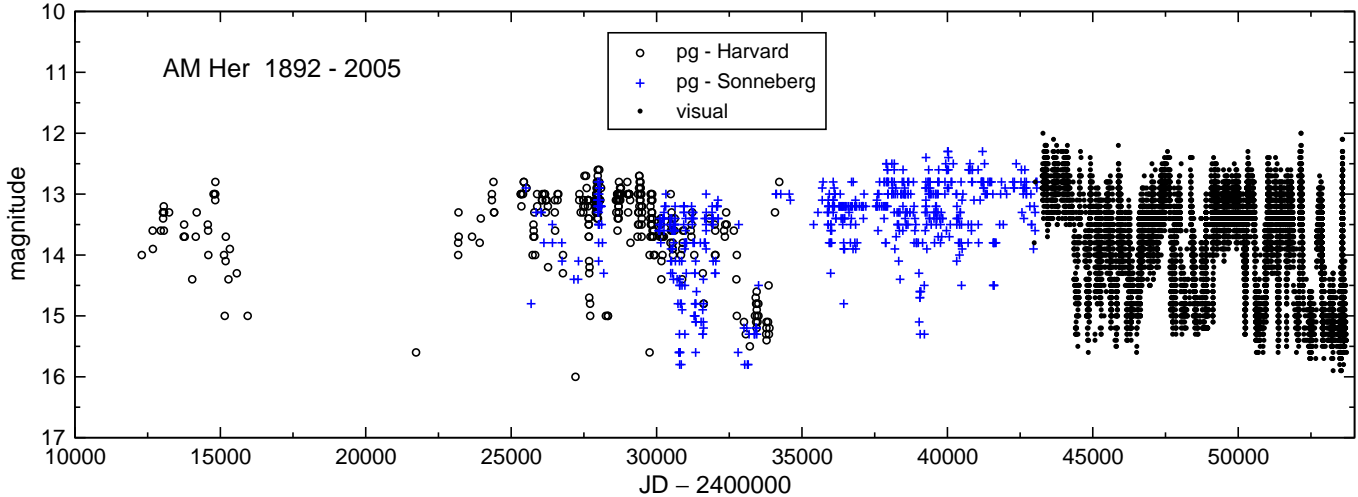}
\end{center}
\caption{The combined photographic and visual light curve of the system AM~Her 
   between 1892 and 2005 (negative detections omitted). }
\label{all}
\end{figure*} 


Our study is based on three different datasets, which allowed the reconstruction of
AM~Her's light curve for most of the last 115 years: {\it (i)} photographic
magnitudes determined by the second author on 348 blue plates from the Harvard
College Observatory's plate collection; {\it (ii)} 600 photographic magnitudes
measured by  Hudec \& Meinunger (1977) on Sonneberg plates; {\it (iii)} 17,704
visual observations collected by the American Association of Variable Star
Observers (AAVSO). The combined light curve is plotted in Fig.~\ref{all}, which
shows that the last $\sim$85 years are covered with almost no long 
(i.e.\ extending many years) gaps.

The earliest observations are those of the Harvard patrol program. The historic
light curve from the Harvard plates was already obtained 
by Feigelson et al.\ (1978); however, those data were never published in tabulated form.
Moreover, we found 378 useful blue plates in the archive, which is 40 more than the
sample analyzed by Feigelson et al.\ (1978). Amongst these, we could detect
AM~Her on 348 plates. Photographic magnitudes were then estimated with reference to
a sequence of 10 nearby comparison stars uniformly covering  the brightness range
from B = 12.6 to B = 16.3. The magnitude values were taken from The Guide Star
Catalog (GSC 2.2, STScI 2001). The brightness estimates of AM~Her were made visually
through a 9$\times$ viewing eyepiece with an estimated photometric accuracy of about
$\pm$0.1--0.2 mag. The earliest plate was taken on August 1, 1890, while the first 
positive detection was on July 18, 1892; the last detection was recorded by us on July
23, 1952. We ignored the few yellow plates because their temporal coverage was very
poor. Also, a few blue plates from 1976--1977 were omitted from the analysis because
for that period the other two sets provided a full coverage. 

It is worth noting that contrary to what Feigelson et al.\ (1978) found,
our Harvard magnitudes are in agreement with those of Hudec \& Meinunger
(1977). In 23 cases when the Harvard and Sonneberg plates were taken on the same
day, the mean Sonneberg minus Harvard magnitude was 0.08$\pm$0.27 mag; since AM~Her
can show much larger variations over a time-scale of hours, the small mean
difference provides reassurance about the homogeneity of the two photographic datasets.

The third and the most continuous dataset comes from the hundreds of visual
observers of the AAVSO. These observations were begun in mid-1977 and have been
continuous ever since. The first 20 years of the data were analyzed 
by Hessman et al.\ (2000) in terms of mass-transfer variations in AM~Her. The
present set is 7.5 years longer, containing almost 18,000 individual observations.
Since we are interested in the transitions between high and low states, we
calculated 10-day bins. This way, we averaged out rapid brightness fluctuations that
occur on shorter time-scales and appear as an additional ``noise'' in the light
curve.

The photographic data are less useful for determining the statistics
of the high and low states. Compared to the visual observations, the photographic
ones are quite sparse. The mean distance between two consecutive Harvard plates  is
35 days (for the more continuous coverage between JD 2,423,000 and 2,433,000), while
the Sonneberg plates were taken only slightly more frequently (one point in every
30 days). For comparison, the mean distance between two consecutive 10-day bins is
10.9 days, which shows that the duty cycle of the visual data was about 90\% with
the 10-day sampling (and still over 70\% with 3-day sampling). Nevertheless, the
photographic curve can still be used to check the range of brightness fluctuations
in the high state over a time scale of a century. Figure \ref{all} reveals that
the overall brightness distribution did not change significantly in the last 113
years. There are hints for gradual change in the maximum brightness, similarly to 
the trend seen in the visual data, 
but the full $\sim$3 mag peak-to-peak amplitude was quite stable.
It is also remarkable how similar the photographic and visual magnitudes were. From
the mean values of the maximum and minimum brightnesses we could not infer any
measurable zero-point shift (greater than 0.1--0.2 mag).

\begin{figure}
\begin{center}  
\leavevmode
\includegraphics[width=8.5cm]{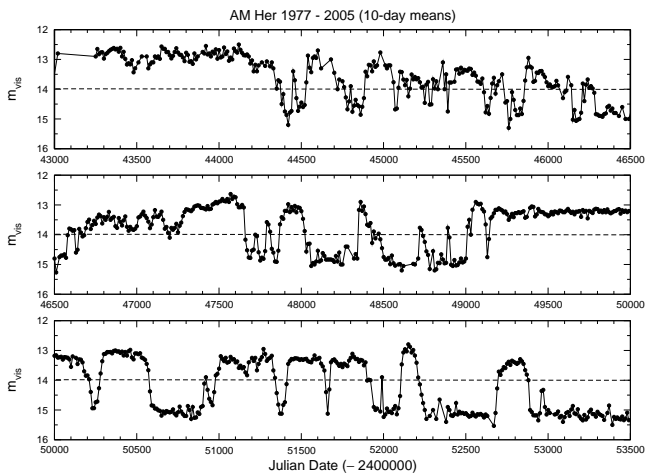}
\end{center}
\caption{The binned AAVSO light curve. The dashed lines 
   show the adopted boundary between high and low state at $m_{\rm vis}=14.0$ mag.}
\label{panels}
\end{figure}

Figure \ref{panels} shows the 10-day binned visual light curve, which is dominated
by the transitions between high and low states. To measure the time-spans of these, we
defined a boundary between them at $m_{\rm vis}=14.0$ mag 
(dashed line in Fig.\ \ref{panels}). 
This way, we could include the short-lived (10--30 days) and 
fainter (13.5--14.0 mag) maxima among the high states. We then measured the
durations of high and low states as the time-spans of being brighter or fainter
than 14.0 mag. Since the typical fading or rising times ranged from 10 to 30 days,
this simplification has only slightly affected the measured lengths, ranging from
10 to 1400 days and 10 to 700 days for the high and low states, respectively.  


\section{Results}  
  
\subsection{Distribution of high and low states and duration of duty cycles}  

\begin{figure}
\begin{center}  
\leavevmode
\includegraphics[width=8.5cm]{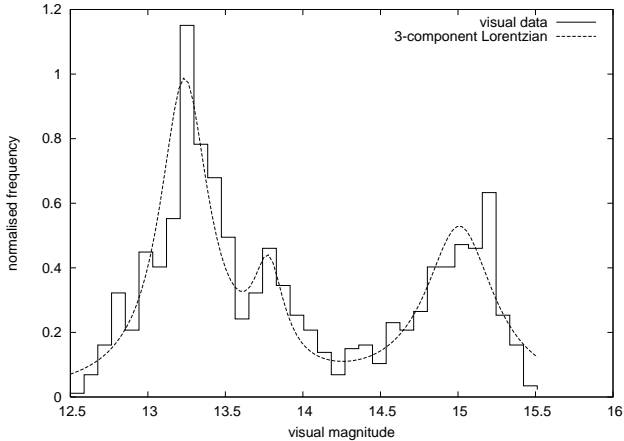}
\end{center}
\caption{The histogram of the binned AAVSO light curve with a three-component Lorentzian fit.}
\label{vishist}
\end{figure}

The normalized histogram of the binned visual light curve (Fig.~\ref{vishist})
clearly shows the two main states of the star. Furthermore, there is a smaller
third peak centered at 13.7 mag, which corresponds to the longer periods of fainter
maxima in the top panel of Fig.~\ref{panels} and the short/faint high states in
the middle and bottom panels of Fig.~\ref{panels}. The integral under the curve
from 12.5 to 14.0 mags gives a measure of the total duration of high states, which
is 62\%, when adding up the area under the steps of the histogram.

\begin{table}
\caption{Best fit Gaussian ($f(x) = C_1\ e^{-(x-\mu)^2/2\sigma^2}$)
and Lorentzian ($f(x) = C_2\ [1+(({x-\mu})/{\Gamma})^2]^{-1}$) 
functions to the histogram of the binned visual light curve.}
\label{fits}
\centering  
\begin{tabular}{|l|c|r|c|}        
\hline                
function & parameters & & rms\\
\hline                       
2 Gaussians &  & & 0.140 \\
 &$\mu_1$     & 13.29$\pm$0.04 & \\
 &$\sigma_1$ & 0.36$\pm$0.04 & \\
 &$\mu_2$ &  14.97$\pm$0.05 & \\
 &$\sigma_2$ & 0.29$\pm$0.05& \\
\hline             
3 Gaussians &  & &0.122    \\
 &$\mu_1$     & 13.23$\pm$0.03 & \\
 &$\sigma_1$ & 0.23$\pm$0.03 & \\
 &$\mu_2$ &  14.97$\pm$0.04 & \\
 &$\sigma_2$ & 0.28$\pm$0.04& \\
 &$\mu_3$ & 13.84$\pm$0.07 &  \\
 &$\sigma_3$ & 0.17$\pm$0.06 &   \\
\hline 
2 Lorentzians &  &  &0.116     \\
 &$\mu_1$     & 13.25$\pm$0.02 & \\
 &$\Gamma_1$ & 0.26$\pm$0.03 & \\
 &$\mu_2$ &  15.01$\pm$0.04 & \\
 &$\Gamma_2$ & 0.28$\pm$0.07& \\
\hline
3 Lorentzians &  & &0.101 \\
 &$\mu_1$     & 13.23$\pm$0.02 & \\
 &$\Gamma_1$ & 0.19$\pm$0.03 & \\
 &$\mu_2$ &  15.01$\pm$0.03 & \\
 &$\Gamma_2$ & 0.27$\pm$0.05& \\
 &$\mu_3$ & 13.78$\pm$0.04 &  \\
 &$\Gamma_3$ & 0.12$\pm$0.07 &   \\
\hline                    
\end{tabular}
\end{table}

To improve the duration calculation, we tried to find a good analytic fit to the
histogram. The best results were given by two- and three-component Gaussian
and Lorentzian functions. Of these, the three-component Lorentzian profile yielded
the smallest residual mean scatter. In Table\ \ref{fits} we give the best fit
parameters of the four functions and corresponding rms values. 
The Lorentzian fits give $\sim$20\% smaller rms, which reflects the observation that the peaks
in the histogram are rather sharp. The integral of the analytic fit gives 62.8\% as
the total duration of the high state, which we adopt as the best estimate of this
parameter with less than 1\% uncertainty (but, of course, subject to systematic bias
arising from the definiton of the high state). 


\begin{figure}
\begin{center}  
\leavevmode
\includegraphics[width=8.5cm]{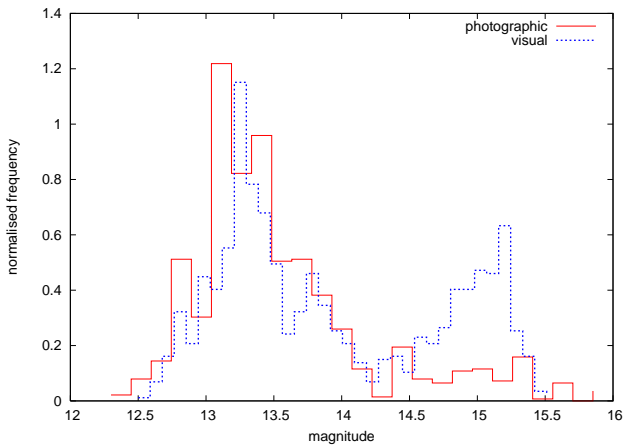}
\end{center}
\caption{A comparison of the histograms of the photographic and visual data. Note
the lack of faint photographic observations.}
\label{pgvis}
\end{figure}

We also calculated the histogram of the photographic light curve. For this, we
combined the Harvard and Sonneberg magnitudes, as there is no systematic difference
between them. In Fig.~\ref{pgvis} we compare it to the histogram of the visual
light curve. Note that the magnitude bins in the histograms are slightly 
different, in order to avoid oversampling of the sparser photographic curve. The
close agreement of the brighter peaks indicates the stability of the full magnitude
range of the high state over the last century. The secondary peak at $\sim$13.7~mag is
not resolved, but the broader shoulder of the main peak suggests the presence of
that distinct feature in the early data. The low state, on the other hand, is
heavily affected by the magnitude limits of the patrol plates. Nevertheless, the
extension of the fainter peak is very similar in both histograms, so that it is very
likely that the statistics of the low state also stayed very similar throughout
the 20th century.


\begin{figure}
\begin{center}  
\leavevmode
\includegraphics[width=8.5cm]{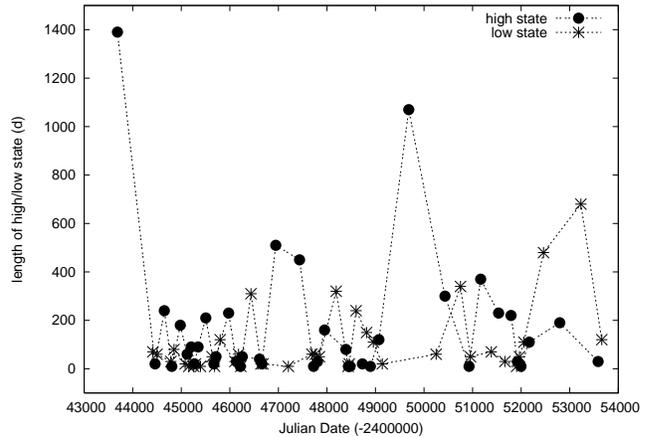}
\end{center}
\caption{Time variations of the duration of high/low states.}
\label{highlow}
\end{figure}

We find a remarkable behavior in the duration of the high and low states. 
In Fig.~\ref{highlow} we plot the time-spans of the two states as a function of time, 
  where the durations are assigned to the mid-times of each state. 
There seems to be a well-defined recurrence in both states, 
  in the sense that alternating long and short states occur over time. 
If this pattern is not due to statistical coincidence, 
  the system has retained certain memory of the previous states and the transitions.    
  
  
\subsection{Comparison with results from X-ray surveys} 

In the {\it ROSAT} X-ray survey  of 28 polars, 16  were in  the low state, and   
  in the {\it XMM-Newton} X-ray survey of the 37 systems, 21  were in the low state.  
Bayesian analyses give a probability of 0.585 for the polars in the low state,   
  with the 90\% confident intervals being (0.342, 0.863),  for the {\it ROSAT} sample, 
  and a probability of 0.442,  with the 90\% confident interval being (0.258, 0.734),  
   for the {\it XMM-Newton} sample (Ramsay et al.\ 2004).   
Moreover,  there were  no significant differences in the global distributions of orbital periods 
   among the high-state and the low-state systems 
   in either the {\it ROSAT} or {\it XMM-Newton} samples. 
The high-state and the low-state systems were not statistically biased 
    toward long or short  orbital periods,   
    and there was no evidence of period clumping of high- and low-state systems.    
The distributions of the high-state and the low-state systems 
   were consistent with random orders,      
   when ranked according to their orbital periods.   
The findings that 
   half of the sources are in the high state and the other half in the low state,   
   and that the high and low states of the systems are unrelated to the system orbital periods   
   suggest a 50/50 high-low state duty cycle for individual systems.    
   
Our analysis of the AAVSO optical photometric data 
    showed that the system AM Her has spent about 60\% of its time in the high state.   
If polars generally have the same high-low state duty cycle as the system AM Her, 
   we would expect that 3/5 of the systems will be in their high state in a snap-shot survey.      
The inference of the X-ray survey observations of Ramsay et al.\ (2004)    
   that the high/low-state fractional duty cycle of an individual polar is about 50\% 
   is therefore consistent 
   with the observed high-low state duty cycles of the system AM Her in the last 100 years.

  
\section{Models for high/low states}   

\subsection{Variations in optical brightness and mass transfer} 

The total bolometric luminosity of a polar is the sum of 
   its hard X-ray, soft X-ray, UV and optical  luminosities, i.e. \ 
   $L_{\rm bol} = L_{\rm hx} + L_{\rm sx} + L_{\rm uv} + L_{\rm opt}$. 
The hard X-rays are free-free emission from the shock-heated accretion gas 
   at the bottom region of the accretion column. 
The soft X-rays probably originate from the thermalized atmosphere 
    of the accretion pole bombarded by dense accretion blobs.  
The UV radiation is generally interpreted as reprocessed emission of the X-rays, 
   and its luminosity is roughly proportional to the X-ray luminosity.       
In the high state, the optical emission is dominated 
   by the cyclotron emission from the shock-heated gas near the white-dwarf surface.  
In the low state, where cyclotron emission from the accretion gas is negligible, 
   the optical luminosity is mainly due to the emission of the low-mass mass-donor star.  
From the {\it BeppoSAX}, {\it ROSAT} and {\it EXOSAT} observations, 
   Hessman et al.\ (2000) demonstrated a scaling relation 
   $L_{\rm opt} \propto  L_{\rm bol}^\gamma$ for the system AM Her 
    in a bright (high) state, where mass transfer occurred.     
The index parameter $\gamma \approx 2$ for $V$ magnitude $< 15$.  
  
The accretion luminosity of an accreting white dwarf is  
   $L_{\rm acc} =  -\  {G M_{\rm wd} {\dot M}}/{R_{\rm wd}}$,  
   where $G$ is the gravitational constant,  
   $M_{\rm wd}$ and $R_{\rm wd}$ are the mass and radius of the white dwarf, respectively,  
   and $\dot M$ is the mass transfer rate.  
For polars in a high state, $L_{\rm acc}$ is much higher than the intrinsic luminosity of the two stars. 
Thus, we have $L_{\rm bol} \approx L_{\rm acc}$.  
Polars are Roche-lobe filling systems, with the mass transfer  rate given by 
\begin{equation} 
   {\dot M}  \approx  - A_{\rm L1} c_{\rm s} ~\rho_0  \exp \left[{-\left(\Delta r/H\right)^2} \right]     
\end{equation}   
    (Lubow \& Shu 1975; Papaloizou \& Bath 1975; Meyer \& Meyer-Hofmeister 1983), 
    where 
    $c_{\rm s}$ is the gas sound speed at the L1 point,  
    $\rho_0$ is the characteristic atmospheric density of the companion star, 
    $H$ is the atmospheric scale height  of the companion star, 
    and $\Delta r$ is the difference between the effective Roche-lobe volume radius  
    and the radius of the companion star.  
The effective area of the nozzle of the gas stream 
    at the first Lagrangian (L1) point,  $A_{\rm L1}$, 
    is given by 
\begin{equation} 
   A_{\rm L1} =  \frac{2 \pi c_{\rm s} a^3}{G M_{\rm wd} (1+q) k(q)}  \ , 
\end{equation} 
   where  $a$ is the orbital separation, 
    $q$ is the ratio of the mass of the secondary star to the white dwarf, 
   and $k(q)$ is a function with value $ \approx 7$ 
   (see Hessman et al.\ 2000; Meyer \& Meyer-Hofmeister 1983). 

The visual magnitude $V$ of an astrophysical object is related to the optical luminosity:  
   $V  =  - \log_{10} L_{\rm opt} + \lambda$, 
   where $\lambda$ is a constant determined by the distance,  
   and the bolometric and photometric corrections.    
It follows that for a polar we have  
\begin{equation}  
  V = - \gamma \left[ \log_{10} \rho_0 
       + 2  \log_{10} c_{\rm s} -   \left(\log_{10} e\right) \xi^2  \right]  + \lambda'  \  ,   
\end{equation}       
   where $\xi = \Delta r/H$.  
When the system parameters $\lambda$, $M_{\rm wd}$, $a$ and $q$ are specified, 
   the quantity $\lambda'$ is a `constant'.  
The visual magnitude $V$, however, is an observable quantity (which is a random variable)  
   determined by the indirectly observables   
   $\rho_0$, $c_{\rm s}$, and $\Delta r/H$ (which are also random variables).    
Any variations/fluctuations in these indirectly observable variables 
   would lead to variations/fluctuations in the visual magnitude $V$.   
It follows from Eq.~(3) that  
\begin{equation} 
  \left(\delta V \right)^2 =  \left( \frac{ \gamma} {\ln 10} \right)^2 
     \left[  \left( \frac{\delta \rho_0}{\rho_0}  \right)^2 
        + \left( \frac{2 \delta c_{\rm s}}{c_{\rm s}}   \right)^2  \right] 
        + \left( \gamma  \log_{10}e  \right)^2 \left(2 \xi \delta  \xi \right)^2 \ . 
\end{equation}  
   
The two-Gaussian and two-Lorentzian fits to the optical brightness variations 
    of the AAVSO data of the system AM Her  
    both yield a mean brightness $\mu_1 \approx 13.3$,     
    and  a standard deviation $\sigma_1 \approx 0.3$ (see section 2) for the high state.   
This gives a characteristic high-state brightness variation $\delta V \sim  \sigma_1 \approx 0.3$. 
In order to explain the spread in the optical brightness during the high state,  
   fluctuations of either $(\delta \rho_0 /\rho_0) \sim 0.4$, $(\delta c_{\rm s}/ c_{\rm s}) \sim 0.2$ 
   or $\delta \xi \sim 0.2$ (for $\xi \sim 1$) are required (Eq.~4).
The low-state mean brightness is about $15.0$ and the standard deviation is about 0.05. 
The difference between the optical brightness of the high and low states 
   is $\delta \mu = \mu_2 - \mu_1 \approx 1.7$.  
During the low state the optical luminosity is contributed mainly 
  by the emission from the mass-donor star and the heated white dwarf 
  (see Bonnet-Bidaud et al.\ 2000).      
Therefore we can use the brightness variation to constrain 
   the variations of $\rho_0$, $c_{\rm s}$ or $\xi$.  
   
\subsection{Star-spot model and magnetic-locking model}     
  
The magnetic-spot  model (Livio \& Pringle 1994) attributes  
   the occurrence of the low state to the cessation or reduction 
   of the mass-transfer rate when a magnetic-star spot 
   on the surface of the mass-donor secondary star 
   traverses the inner Lagrangian point. 
If the area covered by the star spot 
   is not a large fraction (say less than 50\%) of the total stellar surface area, 
   the high state is a `default' state, 
   where the mass transfer is uninterupted.      
Unless there is  an underlying mechanism 
   to synchronize the movement of magnetic spots on the stellar surface,   
   the migration of the star spot to the inner Lagrangian point  is random,  
   and the occurrence of low state and the duration of the low state 
   can be modelled by Poisson processes.  
Moreover, one would also expect that brightness (mass-transfer) variations 
   during the high state are Poisson/Gaussian-type fluctuations,   
   provided that the distribution of the hot-spot size 
   is not narrow and resembling a $\delta$-function.  
   
Analysis by Hessman et al.\ (2000) 
   had highlighted some difficulties of the generic star-spot model.   
In addition, the star-spot model has difficulties 
   in explaining why this kind of high-low state phenomenon occurs in all polars 
   but does not occur in other cataclysmic variables (magnetic or non-magnetic) 
   and low-mass X-ray binaries in general, 
   when all these systems have late-type low-mass secondary stars similar to those in polars.  
   
From the long-term brightness variations of the system AM Her,  
  we have quantified three properties of its high state. 
Firstly, there is a spread of brightness ($\delta V \approx 0.3$)  during the high state.  
The value of the spread is robust --   
  it is insensitive to whether Gaussian fits or Lorentzian fits are used;  
  it is also insensitive to whether a two-component fit or a three-component fit is used.  
Secondly, Lorentzian fits to the V-magnitude variations give better statistics than Gaussian fits, 
  in the two-component model as well as the three-component model.  
Thirdly, there is an additional weaker high state 
   (at $V \approx 13.8$, for Lorenztian fits) 
   in addition to the usual high and low states 
   (at $V \approx 13.2$ and $15.1$ respectively, for Lorentzian fits).   
Moreover, the occurrence and duration of high and low states  
  are not well described as Poission processes, 
  and the alternation of the two states 
  appears to show certain memories (see Fig.~\ref{highlow}).    
Given these properties and the uniqueness of the high/low state in polars,  
   we argue that the high/low states of polars 
   are unlikely to be simply caused by random migration of magnetic star spots 
   of the secondary star to the inner Lagrangian point.  

In a low-mass close binary system, 
  the magnetic field of the compact star 
  generally has negligible direct effects on the system orbital dynamics.      
However, in a polar, the magnetic field of the compact primary star  is so strong 
  that it affects the accretion flow and the orbital dynamics of the system.    
This makes polars very unique among all low-mass close binary systems.  
We propose that  the high/low states and their transitions are caused by 
   variations in the magnetic interaction 
   between the white dwarf  and the mass-donor secondary star.  
In other words, the magnetic field of the white dwarf  
  can alter the mass transfer rate. 
In certain contexts the situation resembles that in the RS CVn systems,  
   where the magneto-activities of a component star are greatly influenced 
   by the magnetic field of its companion star 
   (see e.g.\ Uchida \& Sakurai 1985).  
This is in contrast with  the star-spot model, where  
  the magnetic field of the mass-donor plays the central role.

In the canonical model of polars  (Chanmugam \& Wagner 1977), 
  the magnetic stress exerted by the white-dwarf field  
  is strong enough to disrupt the formation of an accretion disk 
  in the white-dwarf Roche lobe. 
Moreover, all components in the binary are magnetically locked 
   into  synchronous rotation with one single period  
   (Wu \& Wickramasinghe 1993, see also Joss et al.\ 1979; 
   Hameury et al.\ 1987; Campbell 1990). 
 
The orbital separation of a polar is given by 
\begin{eqnarray}
 a & = & 4.8 \times 10^9 \left[ \left(\frac{M_{\rm wd}(1+q)}{{\rm M}_\odot }\right) 
   \left( \frac{P_{\rm o}}{3~{\rm hr}}\right)^2 \right]^{1/3}~{\rm cm}  \  ,   
\end{eqnarray}
 where $P_{\rm o}$ is the orbital period.  
The Roche-lobe volume radius of the white dwarf is given by 
\begin{eqnarray} 
  R_{\rm L} & = & a \left(0.500 + 0.227\log_{10} q  \right)^{-1}   
\end{eqnarray}  
  (Plavec \& Kratochvil 1964). 
The surface polar field of the white dwarf in system AM Her 
   is about 20~MG (Schmidt et al.\ 1981; Bonnet-Bidaud et al.\ 2000).  
The mass-donor of the system AM Her is a M4.5$-$M5 star  
  (Latham et al.\ 1981). 
For a white-dwarf of $\approx 0.75~{\rm M}_\odot$,  
    $R_{\rm L} \sim 2.9 \times 10^9~{\rm cm}$. 
If the white-dwarf magnetic field is dipolar, the field strength would be about 140~kG 
   at the inner Lagrangian point.       
The observed mean surface magnetic field strengths of late-type M stars  
  are $\sim 5$~kG (Johns-Krull \& Valenti 1996), 
  and theoretical models predicted field strengths around 10~kG (see Buzasi 1997). 
These values are significantly smaller than the field exerted by the magnetic white dwarf.  
The white-dwarf magnetic field plays a more dominant role 
  in determining the magnetospheric activity in the inner Lagrangian point, 
  where mass transfer from the donor star initiates.    

The surface temperature of the mass-donor M star of the system AM Her is about 3250~K 
  (Bailey et al.\ 1991).  
In the absence of external forces,  
  the photospheric pressure is estimated to be $\sim 10^3 - 10^4$~erg~cm$^{-3}$  
  (see e.g.\ Edwards \& Pringle 1987).   
The energy density of the stellar field of the mass-donor star is $\sim 10^6$~erg~cm$^{-3}$. 
At the inner Lagrangian point,  the energy density of the white-dwarf magnetic field 
  is, however, greater than $10^9$~erg~cm$^{-3}$. 
The atmospheric scale height $H$ is therefore determined by an external force, 
   due to the white-dwarf magnetic field.  
   
If the brightness variations in the high state of the system AM Her are caused 
   by mass transfer variations, 
   it would require 10--20\% variations either in the sound speed $c_{\rm s}$ at the L1 point, 
   the atmosphere base density $\rho_0$, or the quantity $\Delta r/H$ (see section 4.1). 
Although one cannot dogmatically reject the possibility of large amplitude variations 
   in the surface temperature and base density of the mass-donor star, 
   providing a sensible explanation for the cause of such atmospheric variations is not easy.   
The more likely causes would be the variations in $\Delta r/H$.  
The quantity  $\Delta r$ is the difference between the stellar radius and the Roche-lobe volume radius.     
It varies with the secular evolutionary timescale of the binary orbit,   
   which is much longer than the timescales of the brightness fluctuations of the system.  
Variations in the local magnetic field strength or geometrical configuration 
   may result in a change in the atmospheric scale height $H$ of the mass-donor star. 
The local magnetic field at the L1 point 
   is jointly determined by the white-dwarf magnetic moment and its orientation, 
   the stellar atmospheric magnetic field, and the dynamics of the mass transfer flow.  
The linear width of the nozzle at the L1 point is $\approx \sqrt{A_{\rm L1}} \sim 10^9$~cm (Eq.~2), 
   and the local sound speed is $\sim 10^6$~cm~s$^{-1}$ 
   (inferred from the atmospheric temperature of 3250~K).  
It follows that the shortest timescale allowed for the brightness variations is $\sim 10^3$~s, 
   which is the sound crossing time.  
This is the minimum timescale on which the geometry of mass transfer flow is readjusted  
   in response to changes in the magnetic-field configuration at the vicinity of the L1 point. 
Note that as the mass transfer rate has an exponential dependence on  $(\Delta r/H)^2$ (Eq.\ 1),  
   a reduction of $H$ by a factor of 3 would be be enough to quench the mass transfer.  

Although the component stars in polars are expected 
  to be magnetically locked into synchronous rotation,  
  perfect locking is an idealisation rather than a reality. 
Several polars have been found to show asynchronous white-dwarf spin and orbital rotation 
  (e.g.\ BY Cam, Silber et al.\ 1992; Honeycutt \& Katfla 2005). 
Spin-orbit synchronism is preserved only if the torque generated by the accretion flow 
  is counter-balanced by the magnetic torque.   
There are observations that the locations of the accretion spots in certain polars   
   changed in different epochs. 
An interpretation is that spin-orbital asynchronism is continually perturbed,  
  as the accretion flow and hence the accretion torque are not steady   
  (Bailey et al.\ 1993; Wu et al.\ 1994).   

The high-low state and the transition fit very well in this scenario. 
If the magnetic field configuration is sufficiently perturbed, 
   the mass transfer will be suppressed, 
   and the new equilibrium configuration is established for the zero accretion torque situation. 
This also provided a consistent explanation for an intermediate state (Fig.\ 3), 
   which is, in contrast, not easily explained in the frame-work of the conventional star-spot model.   
Moreover, our analysis of the brightness distribution showed   
   Lorentzians rather than Gaussians, especially for the high and the intermediate states, 
   suggesting that the spread of the brightness might not be due to purely random fluctuations 
   but controlled by certain underlying resonant processes. 
A possible explanation for such resonant is that 
  the locking occurs at the minimum energy field configuration 
  in the presence of a multipolar white-dwarf magnetic field 
  (Wickramasinghe \& Wu 1991; Wu \& Wickramasinghe 1993),   
   and any small perturbations will lead to oscillations of the relative orientation of the magnetic field 
   and hence the mass transfer rates and the brightness of the system.    
When the oscillations are dampened, 
   it naturally gives rise to Lorentzian type profiles for the brightness distribution.  

There are two necessary conditions for the operation 
  of the high-low state-transition mechanism described above.  
Firstly, the system must be magnetically locked or quasi-magnetically locked.  
Secondly, the magnetic field of the compact star must overwhelm   
   the magnetic field of the mass-donor star at the L1 point.  
These conditions are satisfied in polars 
  but not in the weaker field mCVs and other binary systems. 
Here, we elaborate this briefly in the case of intermediate polars.  
Consider an intermediate polar with the same orbital parameters and stellar masses  
  as the system AM Her, 
  but a white-dwarf surface polar field strength of $\sim 1$~MG 
  (see Wickramasinghe et al.\ 1991).  
This field is not strong enough to disrupt the entire accretion disk 
   and to enforce spin-orbit synchronism (see Chanmugam \& Wagner 1977). 
Without screening,  the white-dwarf magnetic field is about 7~kG at the L1 point 
  for a dipolar field configuration,      
  and it is weaker if higher-order multi-pole field components are present.  
An accretion disk is a body of fast differentially rotating conducting fluid. 
Its presence causes field screening.    
The actual strength of the white-dwarf magnetic field at the L1 point 
  is significantly lower than 7~kG, 
  and the corresponding energy density is significantly smaller than $10^6$~erg~cm$^{-3}$.  
The magnetic field of the mass-donor M dwarf, however, reaches $\sim 5-10$~kG, 
  giving an energy density of $\sim 10^6$~erg~cm$^{-3}$. 
The stellar field is therefore strong enough  
   to suppress the magnetic harassment at the L1 point by the white dwarf.  
This allows the M star to self-adjust to establish an atmospheric scale height 
   that facilitates the mass transfer process.  
Note that polars could have evolved from long-period intermediate polars, 
  but long-period polars would not evolve to become shorter-period intermediate polars  
  (Chanmugam \& Ray 1984; King et al.\ 1985; Wickramasinghe et al.\ 1991).        
Observations have shown a deficiency of intermediate polars with $P_{\rm o} < 2$~hr 
   in comparison with polars.  
Dipolar magnetic field scales with $r^{-3}$ 
  and high order multi-pole fields have stronger $r$ dependences 
  (where $r$ is the radial distance from the white dwarf).      
Thus, the white-dwarf magnetic field at the L1 point 
   scales with $P_{\rm o}^{-\alpha}$ roughly,  where $\alpha \ge 2$.   
The influence of the white-dwarf magnetic field on the mass-donor M star 
   is generally insignificant for intermediate polars. 
Therefore, intermediate polars are not expected 
  to show high/low states and state transitions like those observed in polars.  
     
In summary, while the star-spot model has highlighted the importance 
   of magnetism in the role for the high and low states and the state transition of polars, 
   analyses have shown that the model has fallen short of satisfactorily explaining 
   the 100-year data of optical brightness variations of the  system AM Her 
   (see Hessman et al.\ 2000). 
Here, we propose an alternative magnetic model. 
The key regulator of the mass transfer is the global magnetic field of the white dwarf,  
   instead of the local magnetic field variations at the surface of the mass-donor star.    
The states correspond to various equilibrium locking configurations  
   in the presence or absence of the accretion torque. 
The Lorentzian distribution of brightness variation can be attributed by dampened oscillations 
   around the equilibrium configuration. 
This scenario is the same frame-work of synchronism of polars 
  as a unique class of interacting binaries. 
It also reconciles with the observations that prominent high/low states occur in all polars,   
   in which the white-dwarf magnetic field is strong enough to affect the dynamics of the whole system, 
    but not in intermediate polars or non-magnetic cataclysmic variables, 
    in which the white-dwarf magnetic field is too weak to influence accretion flow 
    at the distance of the mass-donor star,  
    or to alter the orbital dynamics of the system.


\section{Conclusion}

We investigate the high/low states of the system AM Her using three sources of archival optical data:  
     the photographic plates of the Harvard College Observatory between 1890 and 1953,  
     the visual observations by the American Association of Variable Star Observers 
     between 1978 and 2005, 
     and the published photographic data from Sonneberg plates. 
We determine the statistical probability 
    of the high and low states, their distribution and recurrence behaviors.  
The distribution of the high states is well fitted by a Lorentzian. 
There appears to be an intermediate state. 
The fractional high state duty cycle of the system AM Her is {63\%}, 
   which is consistent with the inference that the duty cycle of the high state of polars is roughly 50\% 
   based on the {\it ROSAT} and {\it XMM-Newton} survey observations.   
The data show no preference of  timescales on which high or low states occur.   
There appears a pattern of long and short duty cycle alternation, 
    suggesting that the state transitions retain certain memories.     
We propose that the high/low states are caused by 
   variations in the magnetic field configuration of the system, 
   and that the magnetic field of the white dwarf plays the key role in regulating 
   the mass flow at the inner Lagrangian point.  
This scenario may overcome certain difficulties unexplained by the star-spot model, 
   where the cessation of mass transfer is caused by magnetic star spots of the mass-donor star 
   traversing the Lagrangian point.

  
\begin{acknowledgements} 
      KW's visit to the University of Sydney was supported 
           by the NSW State Expatriate Researcher Award. 
       LLK has been supported by a University of Sydney Postdoctoral Research
       Fellowship. We sincerely thank variable star observers of the AAVSO whose
       dedicated observations over three decades made this study possible. 
       LLK thanks the kind hospitality of Dr. Arne Henden, the Director of the AAVSO and all
       the staff members at the AAVSO Headquarters (Cambridge, MA) during his visit
       in early 2006. We are also grateful to Dr. Charles Alcock, the Director of
       the Harvard College Observatory, for access to the Photographic Plate
       Collection, and to Alison Doane, the Curator of the Collection, for helpful
       assistance.
\end{acknowledgements} 


\end{document}